\documentstyle[12pt]{article}
\catcode`\@=11
\def\section{\@startsection {section}{1}{0pt}{-3.5ex plus -1ex minus
 -.2ex}{2.3ex plus .2ex}{\raggedright\large\bf}}
\catcode`\@=12
\newskip\humongous \humongous=0pt plus 1000pt minus 1000pt

\newif\ifdtup

\def\oldreffmt#1{\rlap{[#1]} \hbox to 2\parindent{}}

\def\figfmt#1{\rlap{Figure {#1}} \hbox to 1in{}}

\def\bra#1{\left\langle #1\right|}
\def\ket#1{\left| #1\right\rangle}

\def\beq{\begin{equation}}
\def\eeq{\end{equation}}
\def\bea{\begin{eqnarray}}

\def\eea{\end{eqnarray}}
\catcode`\@=11
\def\eqnarray{\stepcounter{equation}\let\@currentlabel=\theequation
\global\@eqnswtrue
\global\@eqcnt\z@\tabskip\@centering\let\\=\@eqncr
\gdef\@@fix{}\def\eqno##1{\gdef\@@fix{##1}}%
$$\halign to \displaywidth\bgroup\@eqnsel\hskip\@centering
  $\displaystyle\tabskip\z@{##}$&\global\@eqcnt\@ne
  \hskip 2\arraycolsep \hfil${##}$\hfil
  &\global\@eqcnt\tw@ \hskip 2\arraycolsep $\displaystyle\tabskip\z@{##}$\hfil
   \tabskip\@centering&\llap{##}\tabskip\z@\cr}

\def\@@eqncr{\let\@tempa\relax
    \ifcase\@eqcnt \def\@tempa{& & &}\or \def\@tempa{& &}
      \else \def\@tempa{&}\fi
     \@tempa \if@eqnsw\@eqnnum\stepcounter{equation}\else\@@fix\gdef\@@fix{}\fi
     \global\@eqnswtrue\global\@eqcnt\z@\cr}

\newcount\hour
\newcount\minute
\newtoks\amorpm
\hour=\time\divide\hour by60
\minute=\time{\multiply\hour by60 \global\advance\minute by-\hour}
\edef\standardtime{{\ifnum\hour<12 \global\amorpm={am}%
	\else\global\amorpm={pm}\advance\hour by-12 \fi
	\ifnum\hour=0 \hour=12 \fi
	\number\hour:\ifnum\minute<10 0\fi\number\minute\the\amorpm}}
\edef\militarytime{\number\hour:\ifnum\minute<10 0\fi\number\minute}

\def\draftlabel#1{{\@bsphack\if@filesw {\let\thepage\relax
   \xdef\@gtempa{\write\@auxout{\string
      \newlabel{#1}{{\@currentlabel}{\thepage}}}}}\@gtempa
   \if@nobreak \ifvmode\nobreak\fi\fi\fi\@esphack}
	\gdef\@eqnlabel{#1}}
\def\@eqnlabel{}
\def\@vacuum{}

\def\marginnote#1{}
\def\draftmarginnote#1{\marginpar{\raggedright\scriptsize\tt#1}}

\def\draft{\oddsidemargin -.5truein
	\def\@oddhead{\sl DRAFT \hfil \today\quad\militarytime}
	\let\@evenhead\@oddhead
	\let\label=\draftlabel
	\let\marginnote=\draftmarginnote
   \def\@eqnnum{(\theequation)\rlap{\kern\marginparsep\tt\@eqnlabel}%
\global\let\@eqnlabel\@vacuum}  }

\catcode`\@=12

\def\lae{\smash{\,\lower .5 ex \hbox{$\,\stackrel<\sim\,$}}}
\def\gae{\smash{\,\lower .5 ex \hbox{$\,\stackrel>\sim\,$}}}

\def\euec{$eu\longrightarrow ec$\/}

\def\eumuc{$eu\longrightarrow \mu c$\/}
\def\epec{$ep\longrightarrow ec~+~any$}
\def\epmu{$ep\longrightarrow \mu~+~any$}
\def\L{{\cal L}}

\def\beq{\begin{equation}}
\def\eeq{\end{equation}}

\def\sutw{${\rm SU}(2)_W$}

\begin{document}
\begin{titlepage}
\begin{center}
\today \hfill    WIS-92/46/MAY-PHYS

\vskip 1 cm

{\large \bf Can HERA See an $eu\longrightarrow ec$ Signal of a Virtual
Leptoquark?}

\vskip 1 cm

Miriam Leurer

\vskip 1 cm

{\em Department of Nuclear Physics\\
The Weizmann Institute\\
Rehovot 76100\\
ISRAEL}

\end{center}

\vskip 1 cm

\begin{abstract}

Virtual leptoquarks could be detected at HERA through some nonstandard effects.
Here we explore the possibility that virtual leptoquarks could be discovered
via \euec{} scattering, assuming integrated luminosity of 200 pb$^{-1}$ and
charm identification efficiency of 1$\%$. We study the implications of low
energy data for the leptoquarks couplings and find that the most relevant bound
for the HERA cross sections comes from inclusive $c\longrightarrow
e^+e^-~+~any$. This bound implies that the \euec{} cross sections for virtual
leptoquarks are just too small for observation of the signal. With an
improvement by a factor of $\sim2$ on the luminosity or on charm identification
it could be possible to see  virtual leptoquarks with {\it maximum couplings}
up to $\sim 1.5-2$ TeV. However, the prospects for discovering the  virtual
particles if their couplings are somewhat below present bounds are very dim. We
point out that this cross section could be very large for leptoquarks lighter
than HERA's kinematical limit, and if such a leptoquark is discovered we
recommend
searching for a possible \euec{} signal. Our results may also serve as an
update on the maximum cross sections for leptoquark mediated \eumuc{}
scattering.

\end{abstract}
\newpage

\end{titlepage}

It is well known that HERA is an ideal machine for the discovery of low lying
leptoquarks. Such particles, if their mass lies below HERA's kinematical limit
and if their coupling to fermions are not particularly small, are expected
to manifest themselves as peaks in the $x$ distribution of the $ep$ cross
section.

To extend the leptoquark search at HERA beyond the center of mass energy, one
has to study nonstandard effects that would be induced by such virtual
particles. In the past, the possibility that a virtual leptoquark could be
discovered if it induced \eumuc{} scattering has been studied\cite{Bigi}. This
process would look at HERA as \epmu, and, since the muon signal is so
prominent, it could enable one to penetrate the TeV scale. In this paper we
will
study the process \euec. Its signature is not as prominent --- it will look at
HERA as \epec{} and to be able to distinguish such a signal one will need to
identify the charm quark.

The best charm identification method available now to the ZEUS collaboration at
HERA is observing a charged $D^*$ through the decay chain
$D^{*+}\longrightarrow D^0\pi^{+}\longrightarrow
K^{-}\pi^{+}\pi^{+}$\cite{ZEUS}. Unfortunately, the efficiency is low. Less
than $50\%$ of the charm quarks will hadronize to a charged $D^*$; the
branching ratio of the first decay in the chain is $55\%$, and of the second
$3.7\%$\cite{pdg}. So, even before taking into account the deficiencies of the
detector, the efficiency cannot exceed $1\%$. Assuming an integrated luminosity
of 200pb$^{-1}$  we will therefore request that the \euec{} cross section be at
least 1pb.

The paper is organized as follows: First, we review the standard model
backgrounds and the cuts that are necessary to control them \cite{ZEUS}. Then
we consider the possibility that an \euec{} scattering could be induced by
nonstandard physics that is {\it not} leptoquarks, and find that the effects of
such physics are completely negligible. Next we discuss low energy data and the
bounds implied on leptoquarks couplings. We use these bounds to calculate  the
maximally allowed \epec{} cross section as a function of the leptoquark mass.
We explain why our results are also a significant update on the \epmu{} cross
section. Conclusions follow.

The most significant standard model backgrounds originate from scattering of
the electron on charm quarks in the proton sea, and from  photon--gluon fusion
leading to the creation of $c\bar c$ and $b\bar b$ pairs.  Both backgrounds can
be significantly suppressed by cuts on $x$ ($x>x_{min}$) and $t$
($|t|>Q^2_{min}$) \cite{ZEUS}, while leptoquark signals are not much affected
by these cuts \cite{Zerwas}. We also note that \euec{} scattering is allowed in
the standard model at one loop, but the cross section is suppressed both by a
loop suppression factor, $(\alpha_W/4\pi)^2$, and by a GIM suppression
factor, $|\sum_iV_{ui}V_{ci}^*m_i^2/M_W^2|^2$ (here $i=d,s$ or $b$, and $V$ is
the Cabibbo-Kobayashi-Maskawa matrix). These suppressions make the standard
model
\euec{} cross section far too small to have any effect in HERA. We therefore
conclude that standard model backgrounds can be controlled by appropriate cuts
on $x$ and $t$.

Addressing the question as to whether \euec{} scattering at HERA could be
induced  by nonstandard physics other than leptoquarks, we will make the
simplifying assumption that such a nonstandard process occurs at tree level and
is mediated by a scalar or vector boson in the $t$, $s$ or $u$ channel. A boson
in the $t$ channel is neutral, and induces FCNC in the quark sector. Such a
boson could even be the Z itself, with some nonstandard couplings. Bosons in
the $s$ or $u$ channels carry $1/3$ or $5/3$ units of electromagnetic charge
and are leptoquarks. Let us investigate the $t$-channel bosons. Since they
induce FCNC in the quark sector there are strong bounds on their couplings. The
experimental bound on $D^0-\bar D^0$ mixing \cite{pdg} implies that
\beq
\frac{(g^q)^2}{M^2}\lae 10^{-7}G_F\;,
\label{FCNCq}
\eeq
where $g^q$ is the flavour changing coupling to the quarks and $M$ is the boson
mass. The coupling to the electron is certainly bounded by
\beq
\frac{(g^e)^2}{M^2}\le \frac{G_F}{\sqrt2}\;.
\label{FCNCe}
\eeq
The bounds (\ref{FCNCq}) and (\ref{FCNCe}) imply
that the \euec{} cross section at HERA will be $O(10^{-6})$pb that is, there
will be no events. We therefore conclude that an \euec{} scattering, if seen,
must be induced by leptoquarks.

When discussing the leptoquarks, we will, for convenience, refer to the charge
1/3 particles. All the bounds on the coupling constants apply to the charge
5/3 particles as well (when interchanging quarks and antiquarks), and the
final results --- maximum cross sections in HERA --- will be presented
separately for the two kinds of particles.

We start by writing down the most general interaction Lagrangian
for the vector and scalar leptoquark:
\beq
\L_{\rm int}    = \left[\vphantom{P^A_B}
                \bar e^c \gamma^\mu(g^u_{\rm L}{\rm P_L} + g^u_{\rm R}{\rm
P_R}) u
             +  \bar e^c \gamma^\mu(g^c_{\rm L}{\rm P_L} + g^c_{\rm R}{\rm
P_R}) c
                       \vphantom{P^A_B}\right]V_\mu \; ,\label{intv}
\eeq
\beq
\L_{\rm int}     = \left[\vphantom{P^A_B}
                       \bar e^c (g^u_{\rm L}{\rm P_L} + g^u_{\rm R}{\rm P_R})u
                    +  \bar e^c (g^c_{\rm L}{\rm P_L} + g^c_{\rm R}{\rm P_R})c
                       \vphantom{P^A_B}\right]\Phi \; . \label{ints}
\eeq
We did not impose \sutw{} gauge invariance. Generally, since \sutw{} is a
broken symmetry, it does not forbid any of the interaction terms but rather
implies that some other, related interactions exist. For example, $\Phi$ is a
mixture of an \sutw{} singlet and a component of a triplet.  Its interactions
are related to those of the other members of the triplet.  The interactions
related to (\ref{intv}) and (\ref{ints}) by \sutw{} were discussed in
\cite{shanker}, \cite{buchw}, \cite{buchwz} and \cite{Bigi}, and in the
following we will show that they have {\it no} implications for our process.

Next, we write down the \euec{} cross sections for the vector and scalar
leptoquarks:
\begin{eqnarray}
\frac{d\sigma_V}{dt}&=& \frac{1}{16\pi}\frac{1}{(s-M_V^2)^2+M_V^2\Gamma_V^2}
                 \left\{[(| g^u_L|^2| g^c_L|^2+
                                  | g^u_R|^2| g^c_R|^2)]
                                                       (\frac{u}{s})^2
                      +[(| g^u_L|^2| g^c_R|^2+
                                  | g^u_R|^2| g^c_L|^2)
                                                       (\frac{t}{s})^2]
                                \right\}\nonumber\\
                  &\equiv& \frac{1}{16\pi}\frac{1}{(s-M_V^2)^2+M_V^2\Gamma_V^2}
                      [g^4_V(\frac{u}{s})^2+\tilde{g}_V^4(\frac{t}{s})^2]
                      \label{sigv}\;,
\end{eqnarray}

\begin{eqnarray}
\frac{d\sigma_S}{dt}&=& \frac{1}{64\pi}\frac{1}{(s-M_S^2)^2+M_S^2\Gamma_S^2}
                       (| g^u_L|^2+| g^u_R|^2)
                                   (| g^c_L|^2+| g^c_R|^2)\nonumber\\
                  &\equiv& \frac{1}{64\pi}\frac{1}{(s-M_S^2)^2+M_S^2\Gamma_S^2}
                       [g^4_S+\tilde{g}^4_S]
                      \label{sigs}\;.
\end{eqnarray}
We have defined $g$ and  $\tilde{g}$ where
$g^4=|g^u_L|^2|g^c_L|^2+|g^u_R|^2|g^c_R|^2$ and
$\tilde{g}^4=|g^u_L|^2|g^c_R|^2+|g^u_R|^2|g^c_L|^2$. Note that only these two
combinations of the coupling constants are relevant for HERA. $M_V$ and $M_S$
are the masses of the vector and scalar leptoquarks and $\Gamma_V$\/ and
$\Gamma_S$\/ are the widths.

Here we should note that the experimentalists will hunt for charm and anticharm
with equal enthusiasm and efficiency. We therefore always sum the cross
sections of \epec{} and of $ep\longrightarrow e\bar c~+~any$. At the quark
level we are interested in \euec{} and in $e\bar u\longrightarrow e\bar c$.
Looking at fig. 1a and 1b, which describe the two scattering processes for the
charge 1/3 and 5/3 leptoquarks, one notes that the leptoquarks always run in
either the $s$ or the $u$ channel. The $s$ channel cross sections are given by
(\ref{sigv}) and (\ref{sigs}). To get the $u$ channel cross sections from those
formulae write: $\frac{d\sigma}{dt}=\frac{1}{16\pi^2s^2}|{\cal M}|^2$, and
interchange the variables $u$ and $s$ in $|{\cal M}|^2$.  Obviously, for both
scattering processes, the only relevant combinations of coupling constants are
$g$ and $\tilde{g}$.

Our next task is to place low energy bounds on the $g$ and $\tilde{g}$
couplings. In this case, $D^0-\bar D^0$\/ mixing does not give a useful bound.
The mixing now occurs through a box diagram which is, of course, suppressed
being one--loop instead of tree level, and in addition could be suppressed due
to some GIM-like mechanism which could be at work amongst the leptons.

Next we look at bounds coming from $D^0$ decay to $e^+e^-$.
The relevant effective interaction for vector leptoquarks is:
\beq
\L_{eff}=\frac{1}{M_V^2+iM_V\Gamma_V}
\bar u   \gamma^\mu \left[(g^u_{\rm L})^*{\rm P_L}
                      + (g^u_{\rm R})^*{\rm P_R}\right] e^c\;
\bar e^c \gamma_\mu \left[(g^c_{\rm L})  {\rm P_L}
                      + (g^c_{\rm R})  {\rm P_R}\right] c
\label{cueeV}\; .
\eeq
In order to get the $D^0\longrightarrow e^+e^-$ decay rate, we
Fiertz--transform $\L_{eff}$. Then, comparing the result
with the PDG bound \cite{pdg} (B.R.$<1.3\cdot 10^{-4}$) we find:
\beq
E\frac{\tilde{g}^4_V}{M_V^4+M_V^2\Gamma_V^2}
<3\cdot10^{-4}G_F^2 \label{boundv1}\; .
\eeq
where $E>1$ is an enhancement factor: It is the ratio of the
$\bra{\bar D^0}\bar u \gamma_5 c\ket{\vphantom{\bar D^0}0}
\bra{\vphantom{\bar D^0}0}\bar u \gamma_5 c\ket{\vphantom{\bar D^0}D^0}$
to
$\bra{\bar D^0}\bar u \gamma^\mu\gamma_5 c\ket{\vphantom{\bar D^0}0}
\bra{\vphantom{\bar D^0}0}\bar u
\gamma_\mu\gamma_5 c\ket{\vphantom{\bar D^0}D^0}$.
Repeating the same procedure for the scalar leptoquarks we find:
\beq
E\frac{\tilde{g}^4_S}{M_S^4+M_S^2\Gamma_S^2}< 4.8\cdot10^{-3}G_F^2
\; . \label{bounds1}
\eeq

To get a bound on the $g$ couplings, we use the CLEO bound
$B.R.(c\longrightarrow e^+e^-~+~any) <2.2\cdot10^{-3}$ at $90\%$ CL \cite{CLEO}
(see also \cite{ARGUS}). Using the effective Lagrangian (\ref{cueeV}), this
bound implies:
\beq
\frac{g^4_V+\tilde{g}^4_V}{M_V^4+M_V^2\Gamma_V^2}<0.088G_F^2 \; .
\label{boundv2}
\eeq
Similarly, one finds:
\beq
\frac{g^4_S+\tilde{g}^4_S}{M_S^4+M_S^2\Gamma_S^2}<0.18G_F^2 \; .
\label{bounds2}
\eeq
Note that the bounds (\ref{boundv1}) and (\ref{bounds1}) on $\tilde{g}^4$
are much stricter than the bounds (\ref{boundv2}) and (\ref{bounds2})
on the sum $g^4+\tilde{g}^4$. Since we are interested in the case where the
bounds are saturated (so that the HERA cross sections are as large as could
be), the $\tilde{g}$ couplings are negligible. This holds for
the whole leptoquark mass range that is of interest for us (up to
$\sim2-3$TeV).

We now comment on a large list of bounds derived in
\cite{shanker}, \cite{buchw} and \cite{Bigi}. Some of these bounds arise
directly from our interaction Lagrangians (\ref{intv}) and (\ref{ints}),
which induce new contributions to processes and quantities that
are strongly suppressed in the standard model, {\it i.e.} to
$\pi^0\longrightarrow e^+e^-$ decay and to $g-2$ and the electric dipole moment
of the electron. The other bounds arise when one takes into account the
\sutw{} symmetry, which implies the existence of other interactions, related to
our Lagrangians. These extra interactions induce new contributions to nuclear
$\beta$ decay, to $\pi^+\longrightarrow e^+\nu$, $K^+\longrightarrow e^+\nu$,
$K^+\longrightarrow \pi^+\nu\bar \nu$ and $K^0\longrightarrow e^+e^-$ decays.
It
turns out that all these bounds, whether derived directly from our Lagrangians
or by using the \sutw{} symmetry to find related interactions, apply to
combinations of the coupling constants of the form $g^i_Lg^j_L$ or $g^i_Lg^j_R$
(where $i,j = u$ or $c$). We may satisfy all of them by suppressing the LH
couplings $g^u_L$ and $g^c_L$. There is no need to suppress the RH couplings.
In other words, all these bounds may apply to  $\tilde{g}$ (which we anyway
decided to neglect) but not to $g$.

Summarizing the bounds on the leptoquark couplings: There are only
two combinations of the coupling constants that are relevant for the HERA cross
sections, $g$ and $\tilde{g}$. The bounds on $\tilde{g}$ are far
stricter and we therefore neglect terms proportional to these
coupling constants. The {\it only} bounds on the $g$ couplings come from
inclusive $c\longrightarrow e^+e^-~+~any$ decay and they are given in
(\ref{boundv2}) and (\ref{bounds2}).

To be able to use the bounds (\ref{boundv2}) and (\ref{bounds2}) we still need
an estimate for the leptoquarks widths. Clearly, the smaller the widths
the larger the cross sections allowed by the bounds. We do not know
the full width of the leptoquark, but we know its partial width to two decay
channels: $eu$ and $ec$. Since we wish the width to be as small as possible we
will assume that there are no other decay modes. Then, using the interaction
Lagrangians (\ref{intv}) and (\ref{ints}), we calculate the widths:
\beq
\Gamma_V=\frac{1}{24\pi}\left(|g^u_{\rm L}|^2+|g^u_{\rm R}|^2
                                  +|g^c_{\rm L}|^2+|g^c_{\rm R}|^2\right)M_V
\; , \label{gamv}
\eeq
and
\beq
\Gamma_S=\frac{1}{16\pi}\left(|g^u_{\rm L}|^2+|g^u_{\rm R}|^2
                                  +|g^c_{\rm L}|^2+|g^c_{\rm R}|^2\right)M_S
\; . \label{gams}
\eeq
To maximize the cross section of \euec{} scattering, we should make the partial
width of the eu channel equal to that of the ec channel. Then
\beq
\Gamma_V=\frac{1}{12\pi}\sqrt{g_V^4+\tilde{g}_V^4}M_V\; , \label{mingamv}
\eeq
\beq
\Gamma_S=\frac{1}{8\pi}\sqrt{g_S^4+\tilde{g}_S^4}M_S\; . \label{mingams}
\eeq
If we now saturate the bounds, neglect the $\tilde{g}$ couplings and use the
last formulae for the widths, we can express all the quantities
that are relevant to the HERA cross sections as functions of the leptoquark
masses. The maximal $g_V$ and $g_S$ are given by:
\beq
g^4_V=0.088G_F^2M_V^4/(1-0.088G_F^2M_V^4/(12\pi)^2)\;.
\label{gv}
\eeq
\beq
g^4_S=0.18G_F^2M_S^4/(1-0.18G_F^2M_S^4/(8\pi)^2)\;.
\label{gs}
\eeq
A graphical description of the maximal $g^2_V$ and $g^2_S$ is given in fig. 2.
Note that $g^2_V$ reaches $4\pi$ at $M_V\approx 1.85$TeV and $g^2_S$ at
$M_S\approx 1.5$TeV. From these masses up, we do not saturate the bounds
(\ref{boundv2}) and (\ref{bounds2}), but rather fix the maximal coupling
constants at $4\pi$.

Once $g_V$ and $g_S$ are given as functions of the leptoquark masses, we may
substitute them in (\ref{mingamv}) and (\ref{mingams}) and get the widths as
functions of the masses. We then substitute the coupling constants and the
widths into the cross section formulae (\ref{sigv}) and (\ref{sigs}) and get
the
maximum cross sections for each leptoquark mass.

To get the \epec{} cross section we have to convolute the \euec{} cross
section with the structure function of the up quark.
\beq
\frac{d^2\sigma}{dx\;dt}(ep\longrightarrow ec)=
f_u(x,\hat{s})\frac{d\hat{\sigma}}{dt}(eu\longrightarrow ec)\;
\eeq
where $x$ is the fraction of the proton momentum carried by the up quark and
$\hat{s}$ is $x\cdot s$ (with $\sqrt{s}=314\;$GeV being HERA's center of mass
energy). $f_u(x,\hat{s})$ is the up quark structure function and
$\frac{d\hat{\sigma}}{dt}(eu\longrightarrow ec)$ is the
\euec\/ differential cross section when the center of mass energy of the
$eu$ system is $\hat{s}$. The structure functions we use are an approximation
to EHLQ set 2 \cite{EHLQ}.

We also have to take into account the cuts we use to get rid of the standard
model backgrounds. The loosest cuts we employ here are: $x>x_{min}=0.1$ and
$|t|>Q^2_{min}=1000$ GeV$^2$. Under these cuts, the two types of backgrounds
are reduced to 3-4 pb each. Another set of cuts we consider is
$x_{min}=0.2$, $Q^2_{min}=5000$ GeV$^2$. Under these, each background reduces
to $O(0.1)$ pb.

Our results for the charge 1/3 vector and scalar leptoquarks are presented in
fig. 3a. The mass range 200-400 GeV is shown in more detail in fig. 3b. It is
convenient to discuss separately the three mass ranges --- light (below HERA's
kinematical limit), intermediate (above the kinematical limit and up to
$\sim1.85$ GeV for the vectors and $\sim1.5$ GeV for the  scalars) and heavy
leptoquarks:\\ (i) The low lying leptoquarks --- the cross sections here are
large and very enhanced relative to those of the heavier particles. This is
because the leptoquark is really, and not only virtually, created in the
machine. The propagator reaches the pole area and, consequently, the cross
section is strongly enhanced. We therefore recommend that an \euec{} signal be
searched for if the $x$ distribution of the cross section reveals the existence
of a light leptoquark.\\
(ii) Intermediate leptoquarks --- the first, immediate conclusion is that the
cross sections of virtual leptoquarks always drop to the 1 pb level or below
it. Note in particular that the looser cuts are not useful here, since the
corresponding backgrounds, being a few pb each, are considerably larger than
the signal. We therefore use the stricter cuts for which the leptoquark signal
is even smaller.  Virtual leptoquarks will therefore not be discovered via
their possible \euec{} signal, unless the luminosity or charm identification
methods are improved. We remark also that above $\sim$500 GeV, the cross
sections settle into  constant, mass independent, values. This is due to the
fact that for such heavy leptoquarks the propagators that appear in the cross
section formulae are essentially identical to those that appear in the low
energy bounds. Saturation of the bounds then eliminates the mass dependence of
the cross sections.\\
(iii) Heavy leptoquarks --- here the cross sections start dropping as the mass
grows. This is because we do not saturate the bounds but rather fix the
coupling constant at $g^2=4\pi$. The cross sections drop so rapidly that even
if the luminosity or charm identification methods are improved, we do not
expect to penetrate deeply into this region.

Our results for the charge 5/3 leptoquarks are presented in fig. 4. The cross
sections of the charge 5/3 {\it virtual} particles are, within a factor of
$\sim 3$ of those of the corresponding charge 1/3 particles. We conclude that,
like their charge 1/3 counterparts, charged 5/3 virtual leptoquarks will not
be seen via \euec{} scattering at HERA. Considering the possibility that there
will be some improvement on the luminosity or on charm identification, we can
see
by studying figures 3 and 4 that it will become possible to penetrate the
region of leptoquarks with masses up to 1.5--2 TeV if they have {\it maximum
couplings}.  The best prospects are for charge 5/3 vectors, then charge 1/3
scalars, then charge 1/3 vectors and finally charge 5/3 scalars. We
should stress again that the cross sections described in our figures are
calculated with optimistic approach towards the actual values of the leptoquark
widths  and, more significantly, with the {\it maximum} allowed coupling
constants, as drawn in figure 2. The cross sections of the virtual leptoquarks
behave like $g^4$. If $g^2$ is smaller by just a factor of 3 from
the present bound, the \euec{} signal of a virtual particle will
never be seen in HERA.

Note that below HERA's kinematical limit, the cross sections of the charge 5/3
leptoquarks are considerably smaller than those of the charge 1/3 particles.
This is because the cross section in this region is enhanced by the $s$ channel
propagator. In the case of the charge 1/3 leptoquarks the $s$ channel is in
$eu$ scattering, while in the case of the charge 5/3 particles it is in $e\bar
u$ scattering. Since the proton is far richer in up quarks than in up
antiquarks, the $s$-channel enhancement is more significant for the charge 1/3
leptoquarks.

Finally, we wish to remark on the cross sections for a possible \eumuc{}
signal. When this process was studied in \cite{Bigi}, there was no
experimental bound available on the leptoquark couplings. Today, in analogy
to the case discussed in this paper, there are bounds coming from
$D^0\longrightarrow e\mu$ decay and inclusive $c\longrightarrow e\mu~+~any$
decay. Since the bounds on these processes are numerically very similar to the
bounds on $D^0\longrightarrow ee$ and $c\longrightarrow ee~+~any$, we find that
the maximum cross sections calculated in this paper are also relevant for
the \eumuc{} signal and therefore serve as an update on the results of
\cite{Bigi}.

In conclusion, we find that an \euec{} signal will not enable us to see virtual
leptoquarks at HERA, if the luminosity is 200 pb$^{-1}$ and charm
identification efficiency is $1\%$. Some improvement (by factor of $\sim 2$) on
the luminosity or on charm identification may enable us to see virtual, charge
5/3 vector leptoquarks and charge 1/3 scalar leptoquarks up to masses of order
1.5--2 TeV. Further improvement (by another factor of $\sim 2$) may let us see
signals of charge 1/3 vector leptoquarks with similar masses. All this applies
only if the leptoquarks couplings are near their bound. Otherwise, further
improvement on luminosity and charm identification is necessary. Virtual charge
5/3 scalars leptoquarks will probably not be seen via an \euec{} signal. While
the case for virtual leptoquarks seems discouraging, the case for real
leptoquarks (lighter than HERA's kinematical limit) is quite interesting: The
cross sections for \euec, particularly when mediated by a charge 1/3
leptoquarks, could be very large. Therefore, if a real leptoquark is discovered
at HERA via a peak in the $x$ distribution of the cross section, it may well be
worth looking for an \epec{} signal induced by it.

\vskip 2.5cm
\noindent
{\bf Acknowledgements}\\
I thank Uri Karshon and Yehuda Eisenberg for telling
me about their interest in investigating an \euec{} signal at HERA, and for
familiarizing me with various relevant experimental issues. Uri Karshon also
provided me with the estimates for the photon-gluon fusion background.
I am grateful to Arie Shapira for discussions on the experiment and
to Yossi Nir and Shmuel Nussinov for many interesting discussions on the
theory.

\newpage
\noindent {\large \bf Figure Captions}

\noindent{\bf Figure 1a and 1b.} Feynman diagrams for \euec{} scattering
and $e\bar u\longrightarrow e\bar c$ scattering via a charge 1/3 leptoquark
(fig. 1a) and a charge 5/3 leptoquark (fig. 1b).

\noindent{\bf Figure 2.} Maximum allowed values for the leptoquark coupling
constants. The solid line describes $g^2_V$ and the dashed one $g^2_S$.

\noindent{\bf Figures 3a and 3b.} Maximum \epec{} cross sections for charge 1/3
leptoquarks, with the looser set of cuts ($x>0.1$, $|t|>1000$ GeV$^2$), and
with the stricter cuts ($x>0.2$, $t>5000$ GeV$^2$). The solid lines
describe the cross section of the vector and the dashed ones the cross section
of the scalar. The standard model backgrounds to \epec{} scattering amount to a
few pb each for the first set and to somewhat under 0.1 pb each for the second
set.

\noindent{\bf Figure 4.} Maximum \epec{} cross sections for charge 5/3
leptoquarks with the two sets of cuts. Solid lines for the vector, and dashed
ones for the scalar.

\end{document}